\documentclass[12pt,preprint]{aastex}

\usepackage{emulateapj5}

\newcommand{\hj}[1]{\vert \mbox{\boldmath{j}}_{#1}        \vert}

\newcommand{\hjvec}[1]{\hat{\mbox{\boldmath{j}}}_{#1}}

\newcommand{\Lvec} {\mbox{\boldmath{{L}}}}

\newcommand{\hjf}  {\vert \mbox{\boldmath{j}}_{\rm fin}   \vert}

\newcommand{\beq}{\begin{equation}}
\newcommand{\eeq}{\end{equation}}
\newcommand{\ba}{\begin{eqnarray}}
\newcommand{\ea}{\end{eqnarray}}
\def\spose#1{\hbox to 0pt{#1\hss}}
\newcommand{\lta}{\mathrel{\spose{\lower 3pt\hbox{$\mathchar"218$}}
      \raise 2.0pt\hbox{$\mathchar"13C$}}}
\newcommand{\gta}{\mathrel{\spose{\lower 3pt\hbox{$\mathchar"218$}}
      \raise 2.0pt\hbox{$\mathchar"13E$}}}

\shorttitle{Cosmological black hole spin evolution}
\shortauthors{Berti \& Volonteri}

\begin{document}

\title{Cosmological black hole spin evolution by mergers and accretion}

\author{Emanuele Berti\altaffilmark{1} and Marta Volonteri\altaffilmark{2}}

\altaffiltext{1}{Jet Propulsion Laboratory, California Institute of Technology, Pasadena, CA 91109, USA}
\altaffiltext{2}{University of Michigan, Astronomy Department, Ann Arbor, MI
  48109, USA}

\begin{abstract}
  
  Using recent results from numerical relativity simulations of black hole
  mergers, we revisit previous studies of cosmological black hole spin
  evolution. We show that mergers are very unlikely to yield large spins,
  unless alignment of the spins of the merging holes with the orbital angular
  momentum is very efficient. We analyze the spin evolution in three specific
  scenarios: (1) spin evolves only through mergers, (2) spin evolves through
  mergers and prolonged accretion episodes, (3) spin evolves through mergers
  and short-lived (chaotic) accretion episodes. We study how different
  diagnostics can distinguish between these evolutionary scenarios, assessing
  the discriminating power of gravitational-wave measurements and X-ray
  spectroscopy. Gravitational radiation can produce three different types of
  spin measurements, yielding respectively the spins of the two black holes in
  a binary inspiral prior to merger, the spin of the merger remnant (as
  encoded in the ringdown waves), and the spin of ``single'' black holes
  during the extreme mass-ratio inspiral (EMRI) of compact objects. The latter
  spin population is also accessible to iron-line measurements. We compute and
  compare the spin distributions relevant for these different observations. If
  iron-line measurements and gravitational-wave observations of EMRIs only
  yield dimensionless spins $j=J/M^2>0.9$, then prolonged accretion should be
  responsible for spin-up, and chaotic accretion scenarios would be very
  unlikely.  If only a fraction of the whole population of low-redshift black
  holes spins rapidly, spin-alignment during binary mergers (rather than
  prolonged accretion) could be responsible for spin-ups.

\end{abstract}

\keywords{cosmology: theory -- black holes -- gravitational waves -- galaxies: evolution}


\section{INTRODUCTION}
\label{intro}

In general relativity, the properties and dynamics of astrophysical black
holes (BHs) are determined only by their masses and spins. There is
observational evidence that BHs with masses in the range $\sim
10^6-10^9~M_\odot$ exist in the bulges of nearly all local, massive galaxies,
including our own.  BH mass measurements show an almost-linear relation
between the BH mass and the mass of the galactic bulge hosting the BH. Spin
measurements are generally more uncertain
\citep{2005NJPh....7..199N,2007ARA&A..45..441M}, and the cosmological
coevolution of BH masses and spins is a significant open problem in our
understanding of quasars and AGNs
\citep{2004ApJ...602..312G,2005ApJ...620...59S}. 

Massive BH formation scenarios \citep{2003ApJ...582..559V} are important in
the planning of the space-based gravitational-wave detector LISA.
Cosmological BH evolution affects the rates of detectable events
\citep{2006CQGra..23S.785B,2007MNRAS.377.1711S}, data analysis strategies
\citep{2007CQGra..24..551A} and the science performance of the instrument,
including its parameter estimation capabilities. BH spins have a strong impact
on source modeling and parameter estimation, because they induce precession in
the orbits of inspiralling BH binaries and affect the radiation emitted in the
merger and ringdown phases.

\cite{2005ApJ...620...69V} (henceforth paper I) studied the distribution of
massive BH spins and its evolution with cosmic time under the combined effect
of mergers and accretion in the context of hierarchical galaxy formation
theories. They found that gas accretion affects the spin evolution more than
mergers, and that prolonged accretion efficiently spins holes up to $j\sim 1$.

Until recently, the spin of the BH resulting from generic BH mergers was
estimated by heuristic arguments (see e.g. \cite{2002Sci...297.1310M}).
\cite{2003ApJ...585L.101H} (henceforth HB) estimated the merger remnant's spin
using an extrapolation to comparable-mass binaries of results valid for small
mass ratios, $q=M_2/M_1\ll 1$, showing that mergers typically spin BHs down.
Their calculations become unreliable for ``major'' mergers with $q\sim 1$.
Recent breakthroughs in numerical relativity have provided us with a
quantitative understanding of the spins and recoil velocities resulting from
comparable-mass binary BH mergers: see \cite{2007arXiv0710.1338P} for a
review.  Here we update the analysis of paper I, that used the HB model.  For
the first time, we implement numerical relativity results to compute the BH
spin resulting from generic mergers in hierarchical models of BH formation.
We study how the efficiency of spin alignment with the orbital angular
momentum $\Lvec$ affects upper and lower limits on the remnant's spin by
considering three different scenarios: (1) complete isotropy, (2) efficient
alignment of the spin of the more massive hole with $\Lvec$, so that the
smaller hole orbits in the equatorial plane of the larger, and (3) alignment
of both spins with $\Lvec$.

Spin-up is a natural consequence of prolonged disc-mode accretion: any hole
that has doubled its mass by capturing material with constant angular momentum
axis would end up spinning rapidly
\citep{1970Natur.226...64B,1974ApJ...191..507T}.  However, when the angular
momentum of the accretion disc is misaligned with respect to the direction of
the BH spin, accretion of counter-rotating material can cause a spin-down of
the hole. The orbits in the inner accretion disc co-rotate or counter-rotate
depending on the ratio between the angular momentum of the disc and of the
hole: if the cosine of the inclination angle $\cos \phi<-j_d/(2j)$, where
$j_d$ is the total angular momentum of the accretion disc, then the disc
counter-rotates \citep{2005MNRAS.363...49K}.  Sustained accretion from a
counter-rotating disc spins down a maximally-rotating hole to $j=0$ after the
hole has increased in mass by a factor $\sim \sqrt{3/2}$
\citep{1970Natur.226...64B}. A complete overflip eventually occurs, and then
accretion of co-rotating material acts to spin up the BH again: a $180^\circ$
flip of the spin of an extreme-Kerr hole will occur after the BH triples in
mass. Paper I argued that the lifetime of quasars is long enough that the
innermost regions of accretion discs align with BH spins (possibly through
spin flips), and hence all AGN BHs should have large spins.

The picture of spin evolution via prolonged accretion was questioned by
\cite{2005MNRAS.363...49K}. \cite{2006MNRAS.373L..90K} suggest that accretion
always proceeds via very small (and short) episodes, caused by fragmentation
of the accretion disc where it becomes self-gravitating.  The hole would then
accrete each time an amount of mass corresponding to a tiny fraction of its
own mass, and all episodes would have uncorrelated directions (``chaotic
accretion''). Since counter-rotating material spins BHs down more efficiently
than co-rotating material spins them up, this scenario implies that BH spins
are very small: accretion of randomly oriented droplets of gas would rapidly
spin down any BH that had its spin increased by a ``major'' merger. The only
rapidly spinning holes would then be those which have recently experienced a
merger, and have {\it not} accreted any matter afterwards.
\cite{2008MNRAS.tmp..229K} refined their chaotic accretion scenario by
statistical considerations, concluding that accretion very rapidly adjusts the
hole's spin parameter to average values $j \sim 0.1-0.3$ from any initial
conditions, but with significant fluctuations $\Delta j\sim \pm 0.2$ about
these values. In this paper we compare the prolonged-accretion scenario and
the chaotic accretion scenario, showing that rapidly spinning holes should be
extremely rare if chaotic accretion is the norm.

In Section~\ref{merger} we review numerical simulations of BH mergers and
discuss their predictions for the final spin.  In Section~\ref{evol} we
present our main results on cosmological spin evolution in different merger
scenarios, with and without accretion.

\section{Final spin from binary merger simulations}
\label{merger}

In the last two years there has been enormous progress in the numerical
simulation of BH binaries.  We now know that equal-mass non-spinning binaries
produce a rotating (Kerr) BH with final dimensionless spin parameter
$\hjf\simeq 0.69$.  For unequal-mass non-spinning mergers with mass ratio $q$,
the final spin is well fitted by the sum of two terms: $\hjf \simeq
2\sqrt{3}q/(1+q)^2-2.029q^2/(1+q)^4$. The first term is an extrapolation to
comparable masses of the orbital angular momentum of a particle at the
innermost stable circular orbit of a non-rotating BH, and the second term
accounts for the angular momentum radiated in the final plunge
\citep{2007PhRvD..76f4034B}.

A significant sample of spinning binary BH merger simulations is now
available. Different groups showed that when the initial spins are large and
aligned with the orbital angular momentum $\Lvec$ the binary ``hangs up'',
radiating more energy and angular momentum
\citep{2006PhRvD..74d1501C,2007ApJ...668.1140B,2007PhRvD..76l4002P,2007arXiv0709.2160M}.
Even accounting for the additional angular momentum due to orbital
eccentricity, no violation of cosmic censorship should occur: a binary BH
merger should always result in the formation of a Kerr BH
\citep{2007arXiv0710.3823S}.
%
Simulations have also been carried out for some ``generic'' orientations of
the initial spins, including configurations leading to spin-flips or to
Schwarzschild remnants
\citep{2007ApJ...659L...5C,2007PhRvD..76h4032H,2007PhRvD..76f1502T,2007arXiv0711.1097B}.

Unfortunately, covering the whole parameter space by numerical simulations is
computationally costly and not practical. Semi-analytical models for the Kerr
parameter of the final hole are particularly useful for astrophysical
applications.
One such model, based on point-particle analogies, has been proposed by
\cite{2007arXiv0709.3839B}. Their model can reproduce the final spin computed
by numerical simulations within a few percent, the disagreement being larger
when both spins are large and anti-aligned with $\Lvec$.
Here we use semi-analytical fitting formulas of numerical results derived by
\cite{2007arXiv0712.3541R,2007arXiv0708.3999R,2007arXiv0710.3345R}.
Unfortunately, due to the inaccuracy of large-spin simulations and to errors
in the fitting parameters, the final spin $\hjf$ predicted by the fits can be
slightly larger than the Kerr bound when the initial Kerr parameters of the
holes $\hj{i}$ ($i=1,~2$) are large and aligned with $\Lvec$.  To get rid of
this undesired feature we simply introduce a cutoff on the final spin at
$\hjf=1$. An elegant and potentially very accurate recipe to compute the final
spins has been proposed by \cite{2007arXiv0709.0299B} using symmetry
arguments. Some parameters in their ``spin expansion'' are presently
undetermined, but in principle they can be fixed by a reasonably small number
of dedicated numerical simulations.

The spin expansion model should be able to provide accurate and general
predictions for the final spin in the near future, but the fitting formula of
\cite{2007arXiv0712.3541R} is accurate enough for our preliminary exploration.
Their formula can be used to compute
$\hjf(q\,,\hj{1}\,,\hj{2}\,,\cos\alpha\,,\cos\beta\,,\cos\gamma)$, where
$\cos\alpha=\hjvec{1} \cdot \hjvec{2}$, $\cos\beta =\hjvec{1}
\cdot \hat{\Lvec}$ and $\cos\gamma=\hjvec{2} \cdot \hat{\Lvec}$.

\begin{figure*}[thb]
\includegraphics*[width=0.32\textwidth]{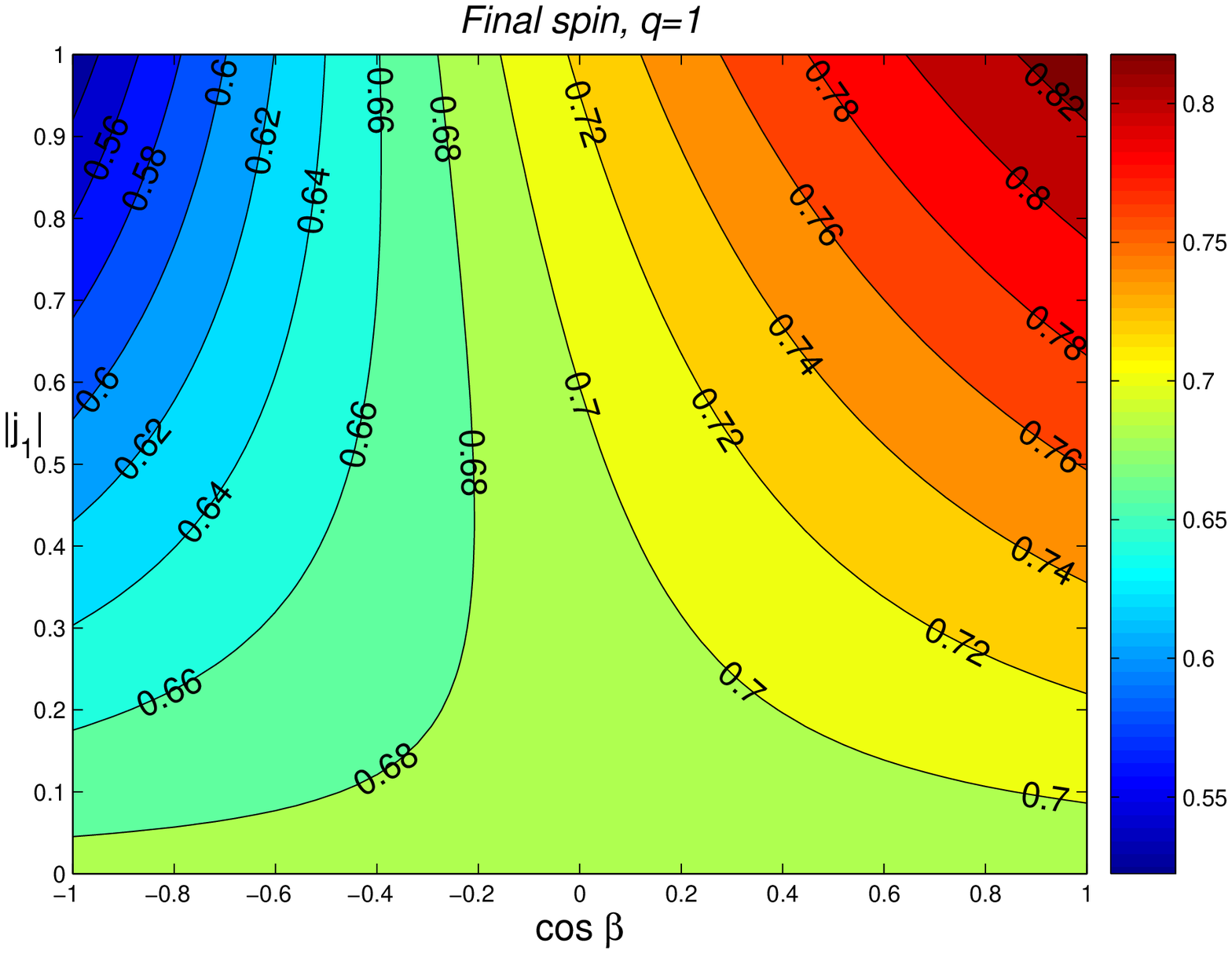}
\hskip 0.2truecm
\includegraphics*[width=0.32\textwidth]{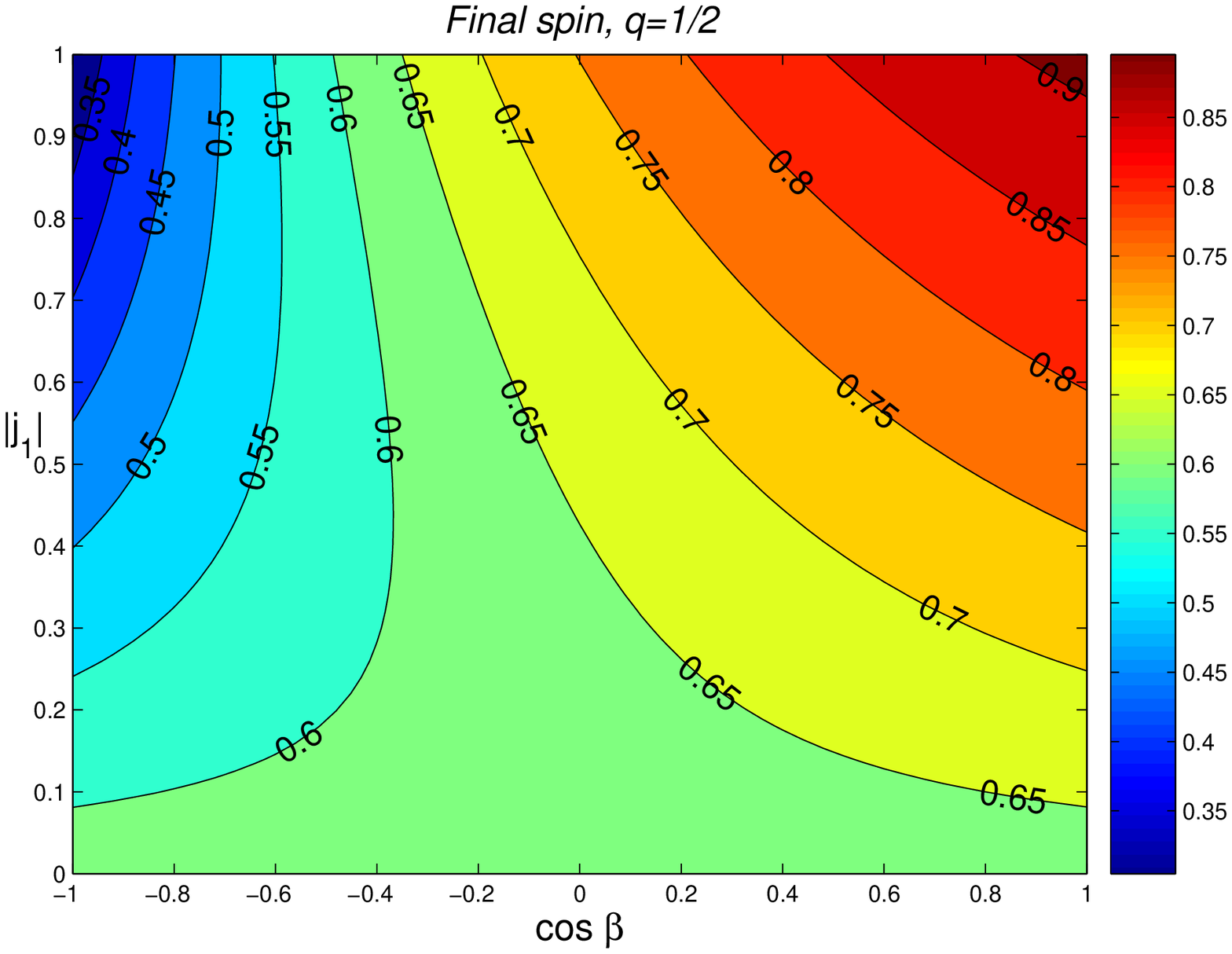}
\hskip 0.2truecm
\includegraphics*[width=0.32\textwidth]{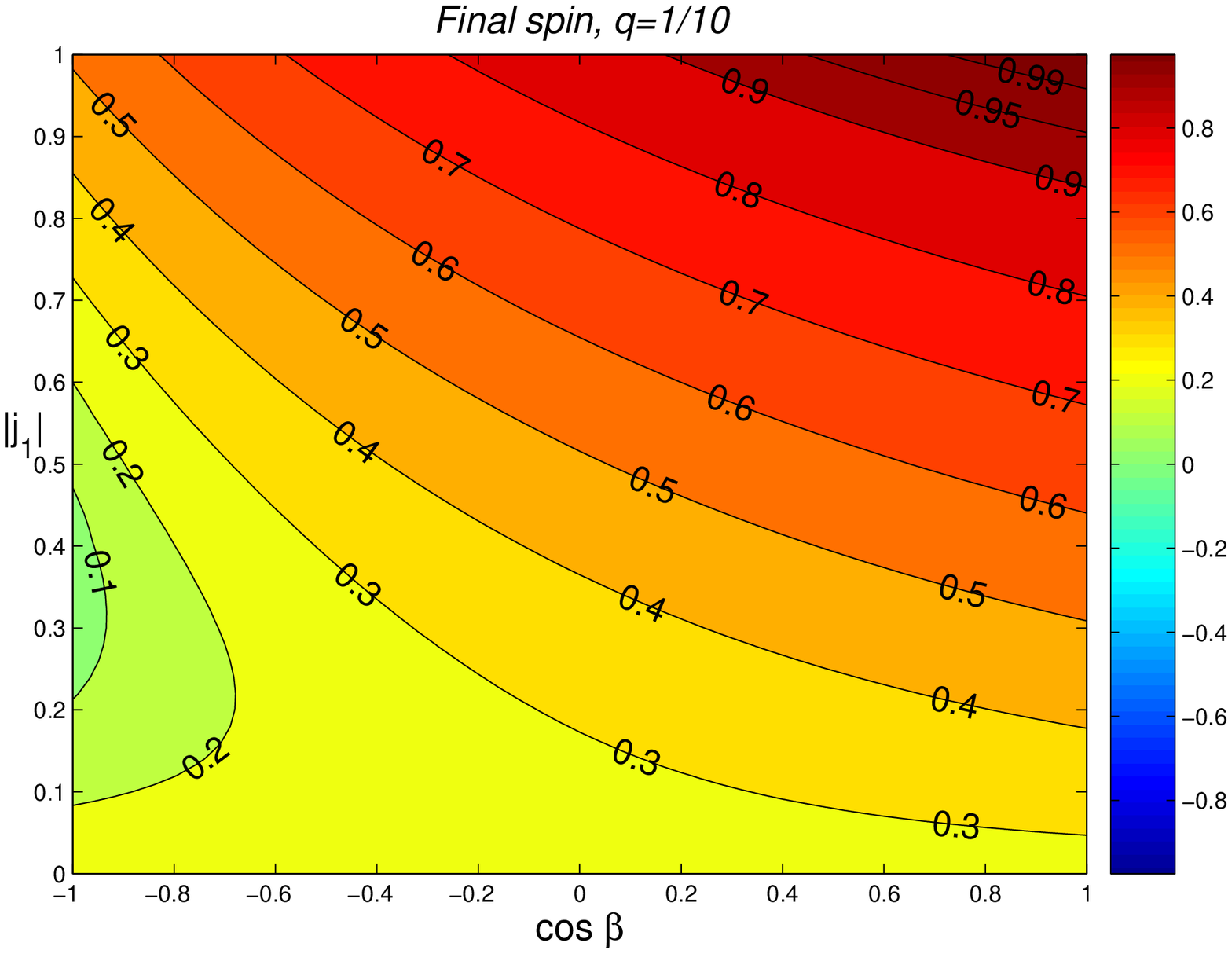}
\caption{\label{fig:HB} Update of Figure~1 in \cite{2003ApJ...585L.101H} .
  Here we assume $\hj{2}=0$ (so that $\cos\alpha$ and $\cos\gamma$ are
  irrelevant) and we show the influence of the orbital inclination
  $\cos\beta$ on the final spin for different values of $\hj{1}$.}
\end{figure*}

As a starting point, for comparison with HB, we computed the final spin
resulting from a merger where the smaller BH is non-spinning ($\hj{2}=0$).
Since $\hj{2}=0$ the angles $\cos\alpha$ and $\cos\gamma$ are irrelevant, and
only $\cos\beta$, the orbital inclination of the smaller hole, matters.
Contour plots of the final spin $\hjf$ in the $(\cos\beta\,,\hj{1})$ plane, as
obtained from the fitting formula, are shown in our Figure~\ref{fig:HB}, that
should be compared with Figure~1 in HB (notice that our angle $\beta$ is
denoted by $\iota$ in their work).
For $q=1/10$ our calculations are in nice agreement with HB: this is a useful
consistency check of our approach in the small mass-ratio limit.  However,
observational and theoretical arguments suggest that the coalescence of
comparable-mass BHs should be rather common \citep{2007ApJ...661L.147B}.  When
$q\sim 1$ and $\hj{2}=0$, we find that the range of variability of the final
spin is sensibly smaller than predicted by HB: $0.3\lesssim \hjf\lesssim 0.9$
for $q=1/2$, and $0.5\lesssim \hjf \lesssim 0.8$ for $q=1$.
%

In Figure~\ref{fig:contours} we drop the assumption $\hj{2}=0$ and we show
contour plots of $\hjf$ in the $(\hj{1}\,,\hj{2})$ plane, for selected values
of $q$. To deal with the angular dependence we consider three different merger
scenarios:

\noindent {\em (1) Isotropic mergers}: for each pair of initial spin
magnitudes $(\hj{1}\,,\hj{2})$ we average over
$(\cos\alpha\,,\cos\beta\,,\cos\gamma)$, assuming isotropy on all three
angles.  This situation should be common in ``dry'' mergers, i.e. when holes
do not accrete during merger, evolving solely via stellar dynamical processes.
In vacuum, post-Newtonian equations predict that spin-orbit resonances will
produce alignment of the spins in a very small region of parameter space
\citep{2004PhRvD..70l4020S,2007ApJ...661L.147B}, so isotropy should be a good
assumption in the absence of accretion discs or gas.  For all values of $q$,
the minimum final spin results from non-spinning mergers with
$\hj{1}=\hj{2}=0$.  Naively we could expect that the average over all angles
$(\cos\alpha\,,\cos\beta\,,\cos\gamma)$ should produce a final spin that is
very close to the spin produced by non-spinning mergers, at least for equal
masses.  As we consider larger mass ratios the larger BH should play a more
important role, so (on average) the final spin should be slightly {\it larger}
than the value predicted by non-spinning mergers. This expectation is
confirmed by the top row of Figure~\ref{fig:contours}.  In general, our
calculations are in agreement with the HB conclusion that isotropic mergers
tend to ``spin-down'' a fast-spinning hole.
%

\noindent
{\em (2) (Anti)aligned mergers}: $\cos\alpha=\pm \cos\beta=\pm \cos\gamma=1$,
so the BH spins are always parallel (aligned or antialigned) with $\Lvec$.  In
the ``wet merger'' scenario proposed by \cite{2007ApJ...661L.147B}, 
alignment (upper signs in $\pm$) should be more likely than antialignment
(lower signs), because the cumulative angular momentum of the accretion disc
is much larger than the angular momentum of the BHs \citep[but
see][]{2005MNRAS.363...49K,2006MNRAS.368.1196L}. If alignment is indeed the
norm, we should focus on the top right quadrant of the contour plots in the
central row. Then the minimum spin is still obtained when $\hj{1}=\hj{2}=0$,
as in the isotropic case. The main difference is that now the maximum spin can
be quite close to one, with a lower limit $\hjf\sim 0.96$ for equal-mass
mergers.

\noindent {\em (3) Equatorial mergers}: this case is intermediate between
cases (1) and (2). Here the smaller BH inspirals in the equatorial plane of
the larger ($\cos\beta=1$), but for each pair $(\hj{1},\hj{2})$ we average
assuming an isotropic distribution in $(\cos\alpha,\cos\gamma)$. Equatorial
mergers could occur if Newtonian dynamical friction in a flattened system
brings the smaller hole into the plane of a central (gaseous or stellar) disc
before it gets into the relativistic regime.  All mergers with $q\geq 1/2$
produce an average final spin $\hjf \leq 0.92$, and $\hjf \leq 0.85$ for
$q=1$. As in case (1), to get large Kerr parameters $\hjf >0.9$ in the absence
of prolonged accretion we must start with large spins {\it and} avoid major
mergers.

For each value of $q$, it is interesting to maximize and minimize the final
spin resulting from a merger in the $(\hj{1},\hj{2})$ plane. The resulting
extrema in the three different scenarios are plotted as functions of $q$ in
Figure~\ref{fig:spinrange}. For each value of $q$ we average over the
individual spin magnitudes $((\hj{1},\hj{2})$. Both $\hj{1}$ and $\hj{2}$ are
assumed to be uniformly distributed between $0$ and $1$. In the isotropic
case, the allowed average values of the final spin range between the solid and
dashed black lines.  The most interesting feature is the appearance of a
funnel at $q\gtrsim 0.1$: major mergers tend to produce BHs with average spins
very close to the value $\hjf \simeq 0.69$ resulting from equal-mass,
non-spinning mergers.  For smaller $q$ the larger BH dominates the dynamics,
and the final spin can be substantially larger or smaller than this
value. Moving from isotropic to equatorial and aligned mergers can produce
slightly larger maximum spins (as indicated by the solid lines of decreasing
thickness), but in all three scenarios the most likely spin resulting from
``major'' mergers is very close to $\hjf \simeq 0.69$. We now explore some
consequences of these results for spin evolution in cosmological BH formation
scenarios.

\section{Cosmological spin evolution}
\label{evol}

In Figure~\ref{fig:mergers-evol} we show histograms of the spin distribution
of {\it merging} BHs in different redshift ranges. For simplicity, BH seeds at
high redshift are assumed to be non-spinning. These plots are useful to
isolate the contribution of mergers to the spin evolution of the whole BH
population, and they are of direct interest for LISA observations of massive
BH inspiral, merger and ringdown
\citep{2005PhRvD..71h4025B,2006PhRvD..73f4030B,2006PhRvD..74l2001L,2007arXiv0712.1144K}.
The three columns in each of the three plots correspond to the merging
scenarios described above: isotropy, efficient alignment of the spins with
$\Lvec$, and equatorial mergers.

In the left plot, spin evolution is due to mergers only and accretion is
ignored.  Consider first isotropic mergers (left column). At high $z$ most BH
seeds have comparable masses, and since mergers with $q\gtrsim 0.1$ produce
remnants with $\hjf\sim 0.7$ (Figure~\ref{fig:spinrange}) the spin
distribution post-merger peaks around this value. Later on, small-$q$ mergers
become more common, and on average (as shown in HB) they tend to spin down the
hole, so the spin distribution flattens out. For aligned mergers (central
column) the evolution is different. At high $z$, mergers of comparable-mass,
non-spinning holes again produce a peak at $\hjf\sim 0.7$. However, if
alignment is efficient mergers of spinning holes have a tendency to spin up
the remnant, and eventually most BHs at small $z$ are rapidly spinning.  In
the equatorial merger scenario (right column) the less massive hole has
marginal impact on the dynamics, and the overall spin distribution is
qualitatively very similar to the case of alignment.

In the central plot spin evolution is due to both mergers and accretion, where
accretion is modeled as in paper I: prograde or retrograde orbits are equally
probable, and spins evolve according to \cite{1970Natur.226...64B}.  If enough
mass is available to the hole, and the disc was initially counter-rotating
with respect to the hole, an overflip can eventually occur.  In this scenario,
accretion-induced spin-up is very efficient. For aligned (equatorial) mergers,
$\gtrsim 90\%$ of merging BHs at $z\lesssim 5$ ($z\lesssim 2$) have $j>0.9$.
If mergers occur isotropically, on average they tend to spin down the holes,
partially counteracting the accretion-induced spin-up. As a result, in this
case the spin distribution post-merger has a long tail extending all the way
down to $j\simeq 0$.

In the right plot we consider the chaotic accretion scenario. Now accretion
happens in short-lived episodes, where the BH increases its mass by 0.1\%, and
prograde or retrograde orbits are equally probable. Mergers {\it per se} are
unlikely to produce fast-spinning remnants, and chaotic accretion is very
efficient in spinning BHs down. Therefore, at all but the highest redshifts
the spin distribution post-merger is roughly uniform in the range $j\in
(0,~0.7-0.8)$, and the chance of mergers involving high-spinning BHs is very
low. LISA measurements of spins $j>0.9$ would strongly favor prolonged
accretion and indicate that alignment is efficient in merger events.
Conversely, measuring values of $j<0.9$ at $z\lesssim 5$ would indicate that
accretion is chaotic (or negligible), or that alignment is not efficient.

For BH spin measurements based on X-ray spectroscopy
\citep{2007ARA&A..45..441M} and for the low-redshift observations of
gravitational waves from compact objects falling into a massive BH (EMRIs),
that could allow very accurate spin measurements \citep{2004PhRvD..69h2005B},
we are interested in the spin distribution of the whole BH population. This
distribution is shown in Figure~\ref{fig:all-evol}.

When we consider the spin evolution {\it under mergers only} (left plot) the
overall spin distribution is bimodal. One peak is around $j\sim 0$ because we
assume that seed BHs are non-spinning, and some BHs never experience mergers.
A second peak is located around $j\sim 0.7$ (for isotropic mergers) and it
extends all the way up to $j\sim 1$ for aligned/equatorial mergers. 

If accretion influences the evolution of BH spins, it dominates over mergers
in determining the spin evolution of the whole BH population, confirming the
results of Paper I. We stress that our accretion model is {\it highly
  oversimplified}, and does not take into account the different evolution of
spins in different galactic environments: here we simply assume that accretion
is triggered {\it always and only} by galaxy mergers. This is indeed a
simplistic assumption, especially for low-redshift faint AGNs: see
\cite{2007ApJ...667..704V} for a detailed analysis of the connection between
spin evolution and galaxy morphology.

Whatever the dominant merger scenario, under prolonged accretion (central
plot) spin-up is very efficient, and a large fraction of all BHs has spin
$j>0.9$.  On average isotropic mergers tend to spin BHs down: if
isotropy is the norm, $\sim 30\%$ of the whole BH population has $j<0.9$, with
a roughly uniform distribution. For aligned and equatorial mergers spin-down
is less efficient, and only $\sim 10\%$ of all BHs have $j<0.9$.

The chaotic accretion scenario (right plot) predicts a completely different
distribution. In this case most BHs ($\sim 50-80 \%$ of the total) have
$j\lesssim 0.1$. When the average spins are so low mergers act to spin BHs up,
and the spin-up is more efficient in the aligned and equatorial cases. This
produces a roughly uniform spin distribution in the range $j\in (0.1,~0.7)$.
Interestingly, the tail of the distribution never extends above $j\sim 0.8$,
even for small redshifts.  The measurement of large BH spins, such as the
value of $j\sim 0.99$ claimed by \cite{2006ApJ...652.1028B} for MCG-06-30-15
\citep[see also][]{2002ApJ...565L..75E,2006ApJ...642L.111W} would presumably
indicate that chaotic accretion is not the norm. In any case, BH spin
measurements should easily distinguish between chaotic and prolonged
accretion.  Both for merging BHs and for the whole BH population, under
chaotic accretion maximal Kerr BHs are extremely unlikely, and the
distribution peaks at $j<0.7$ for all $z<10$.

Suppose that LISA measurements of the spin resulting from a binary merger
yield $j>0.9$. Then it is hard to tell if spin is accretion-independent, but
alignment is efficient (left plot in Figure \ref{fig:mergers-evol}) or if
instead prolonged accretion is responsible for spinning BHs up (central plot).
Luckily, this degeneracy is broken when we consider the whole population. If
accretion is negligible and mergers are not isotropic, only $\lesssim 50\%$ of
all BHs have spins $j>0.9$. If instead spin-up is due to accretion, the
population is largely dominated by spins $j\sim 1$ (compare the left and
central plots in Figure \ref{fig:all-evol}). EMRIs \citep{2004PhRvD..69h2005B}
and X-ray observations of the Fe K$\alpha$ line by planned missions such as
Constellation-X \citep{2007ARA&A..45..441M} will probe the whole population at
low redshifts ($z<2$), while observations of the inspiral
\citep{2006PhRvD..74l2001L} and ringdown \citep{2006PhRvD..73f4030B} of
massive BHs can provide a census of BH spins over a wider range of epochs. A
combination of these techniques (and possibly others) has the potential to
probe unequivocally the physical processes involved in the evolution of BH
spins.

\section{Conclusions}

Understanding how fast BHs spin is as important as understanding how they grow
in mass. The spin and mass evolution of BHs are intimately linked. The
expected spin of a hole depends on whether it gained most of its mass via
mergers or accretion. Conversely, the spin influences how efficiently BHs
accrete mass, determining the mass-to-energy conversion efficiency in
radiatively efficient accretion phases. The spin also determines how much
energy can be extracted from aBH. Assuming that relativistic jets are powered
by rotating BHs through the Blandford-Znajek mechanism, the so-called ``spin
paradigm'' asserts that powerful relativistic jets are produced in AGN with
fast rotating BHs \citep{1990agn..conf..161B}.

Studying the role of BH mergers in the evolution of BH spins has been daunting
until recently.  Due to remarkable advances in numerical relativity, we
finally have a quantitative understanding of the spins resulting from
comparable-mass binary BH mergers \citep[see e.g.][]{2007arXiv0710.1338P}. Here
we use numerical relativity results to investigate plausible scenarios of spin
evolution through mergers and accretion in a cosmological context. The
co-evolution of BH masses and spins is studied self-consistently: mergers and
accretion determine BH spins, which in turn lend to the calculation of the
radiative efficiency during accretion episodes.

We focus on three scenarios for the mass and spin co-evolution: (1) spins
evolve only through mergers, (2) spins evolve through mergers and prolonged
accretion episodes, (3) spins evolve through mergers and short-lived (chaotic)
accretion episodes. If BHs accreted most of their mass through prolonged
disc-mode accretion, by adding material with constant angular momentum axis,
they would end up spinning rapidly. If instead BHs built-up their mass via
short-lived episodes with uncorrelated angular momentum axis, the typical spin
of BHs would be very low.

\begin{figure*}[thb]
\includegraphics*[width=0.32\textwidth]{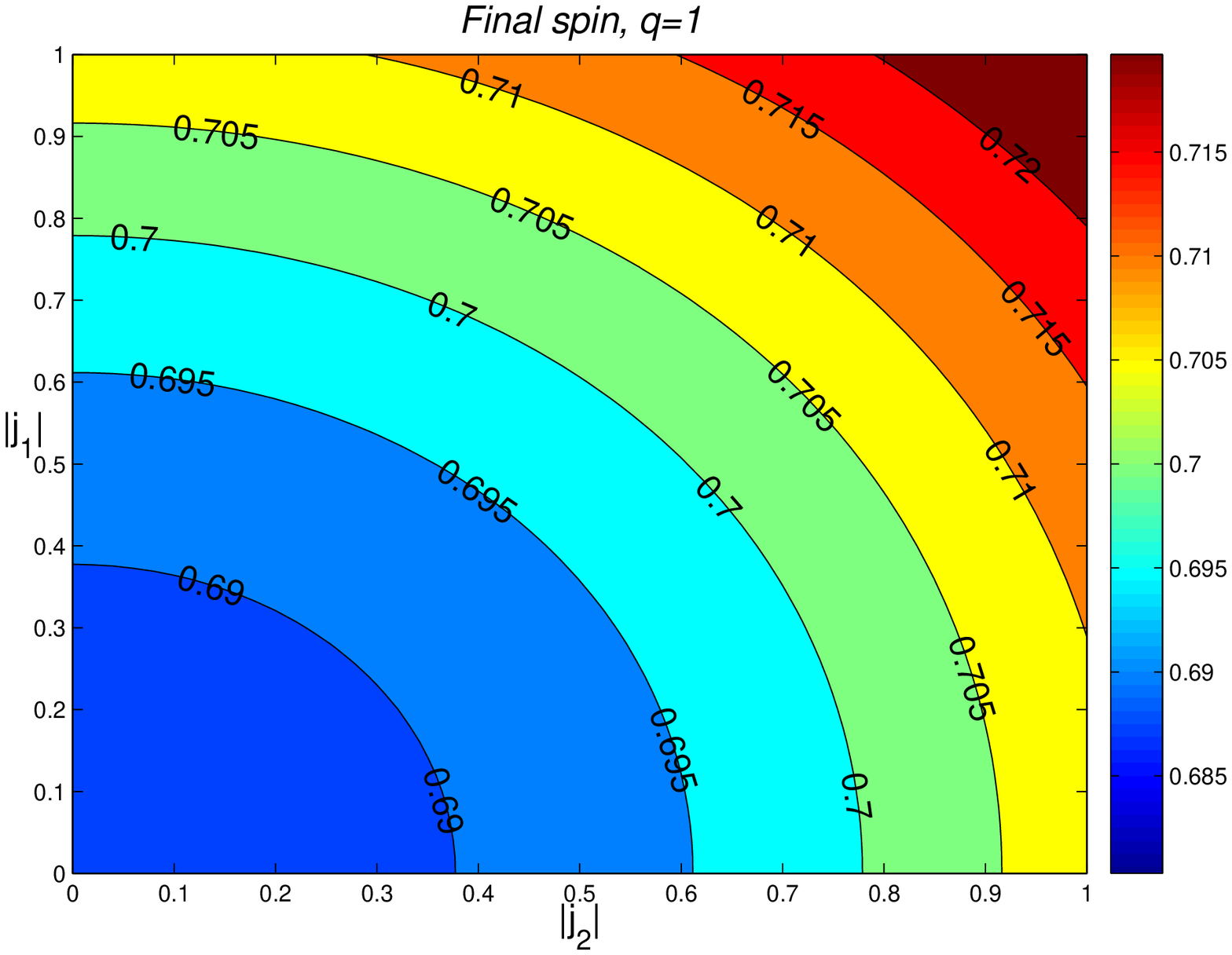}
\includegraphics*[width=0.32\textwidth]{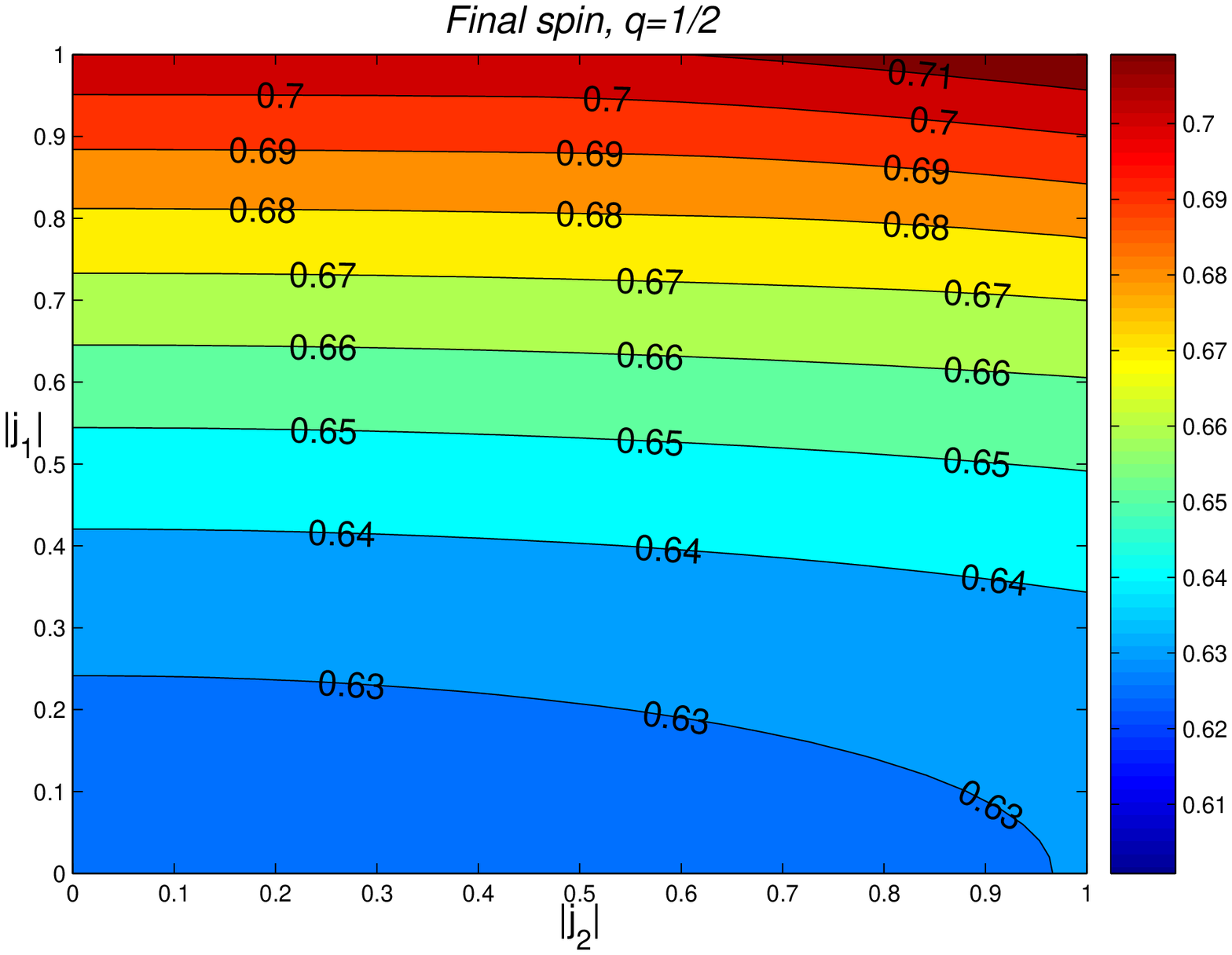}
\includegraphics*[width=0.32\textwidth]{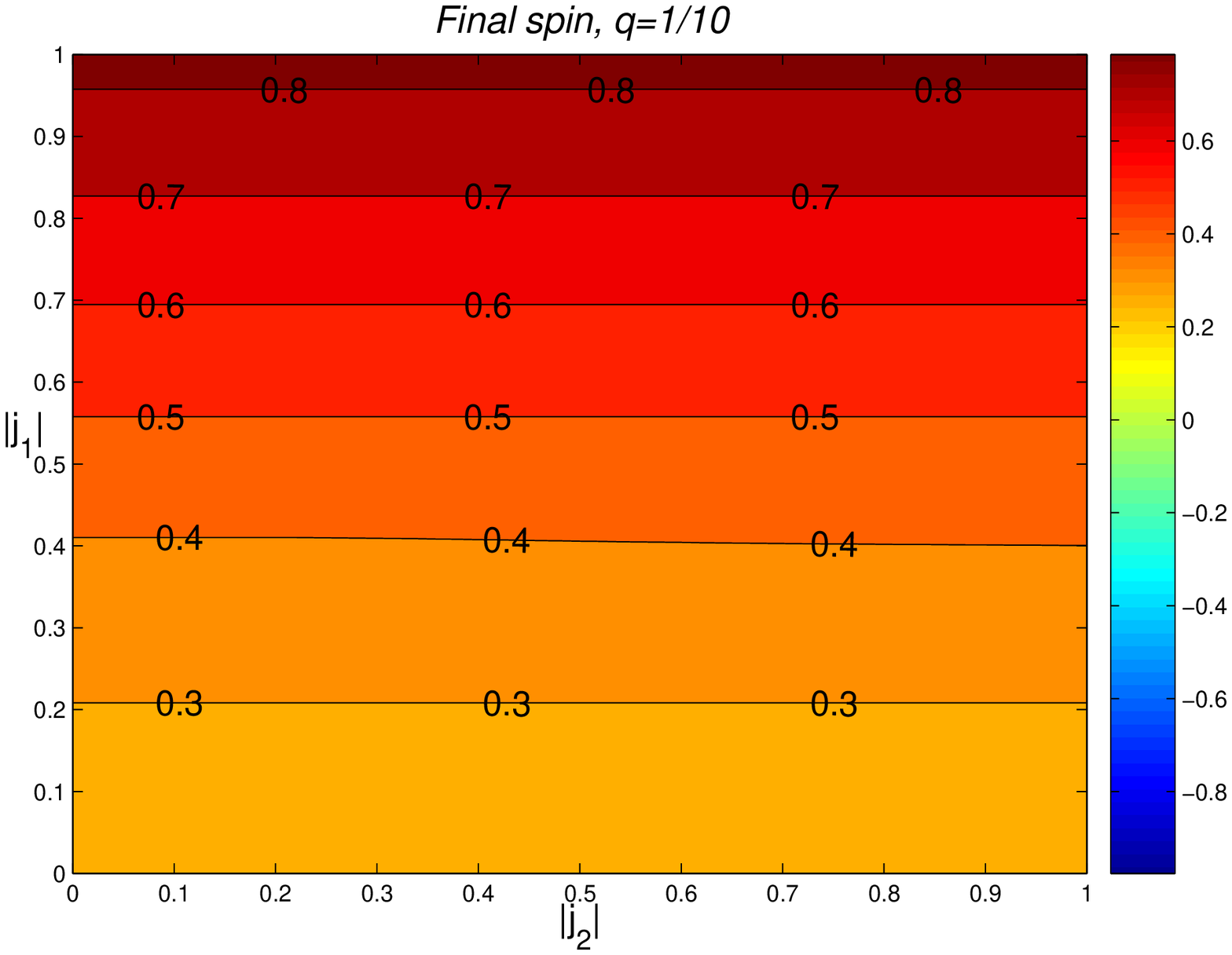}
\includegraphics*[width=0.32\textwidth]{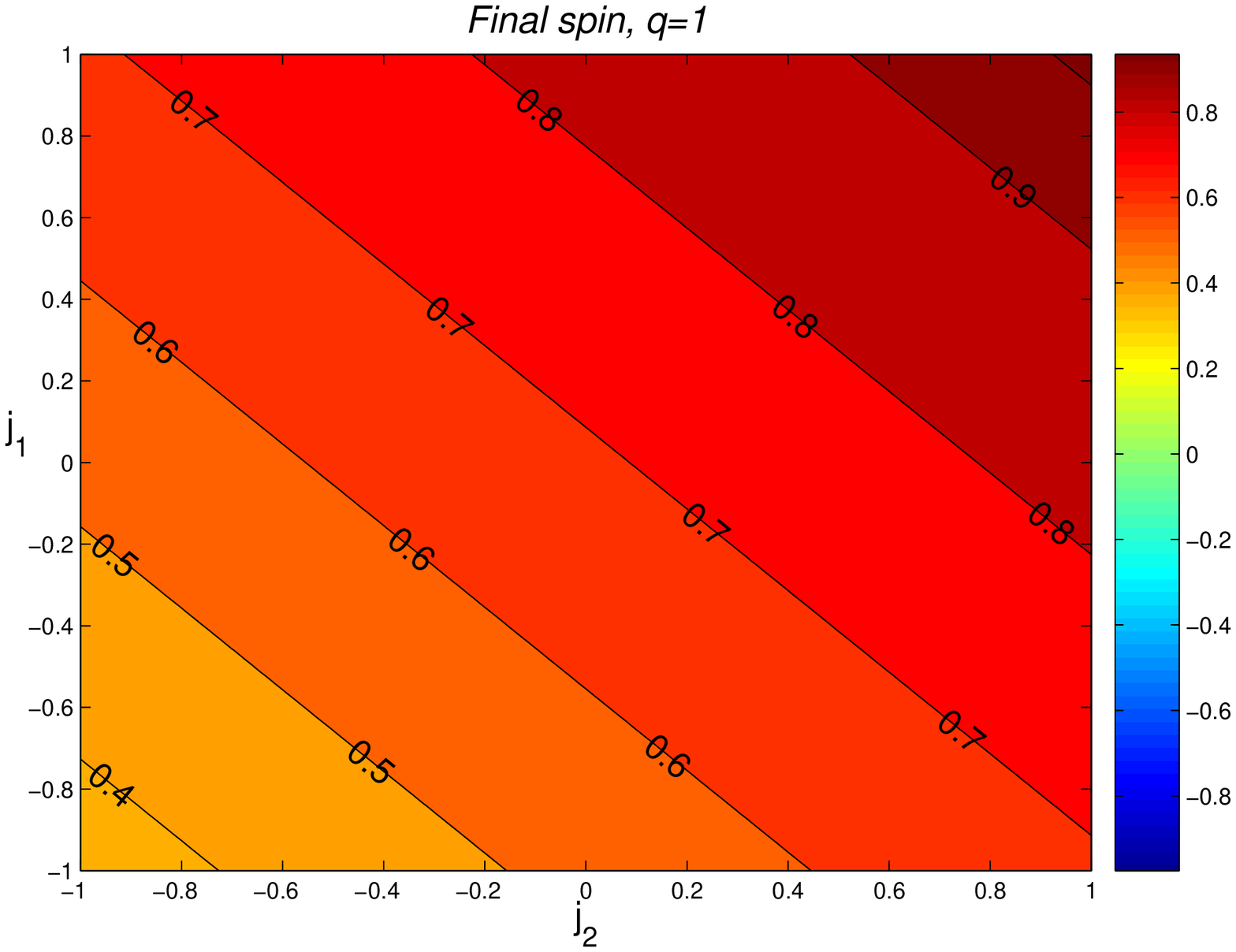}
\includegraphics*[width=0.32\textwidth]{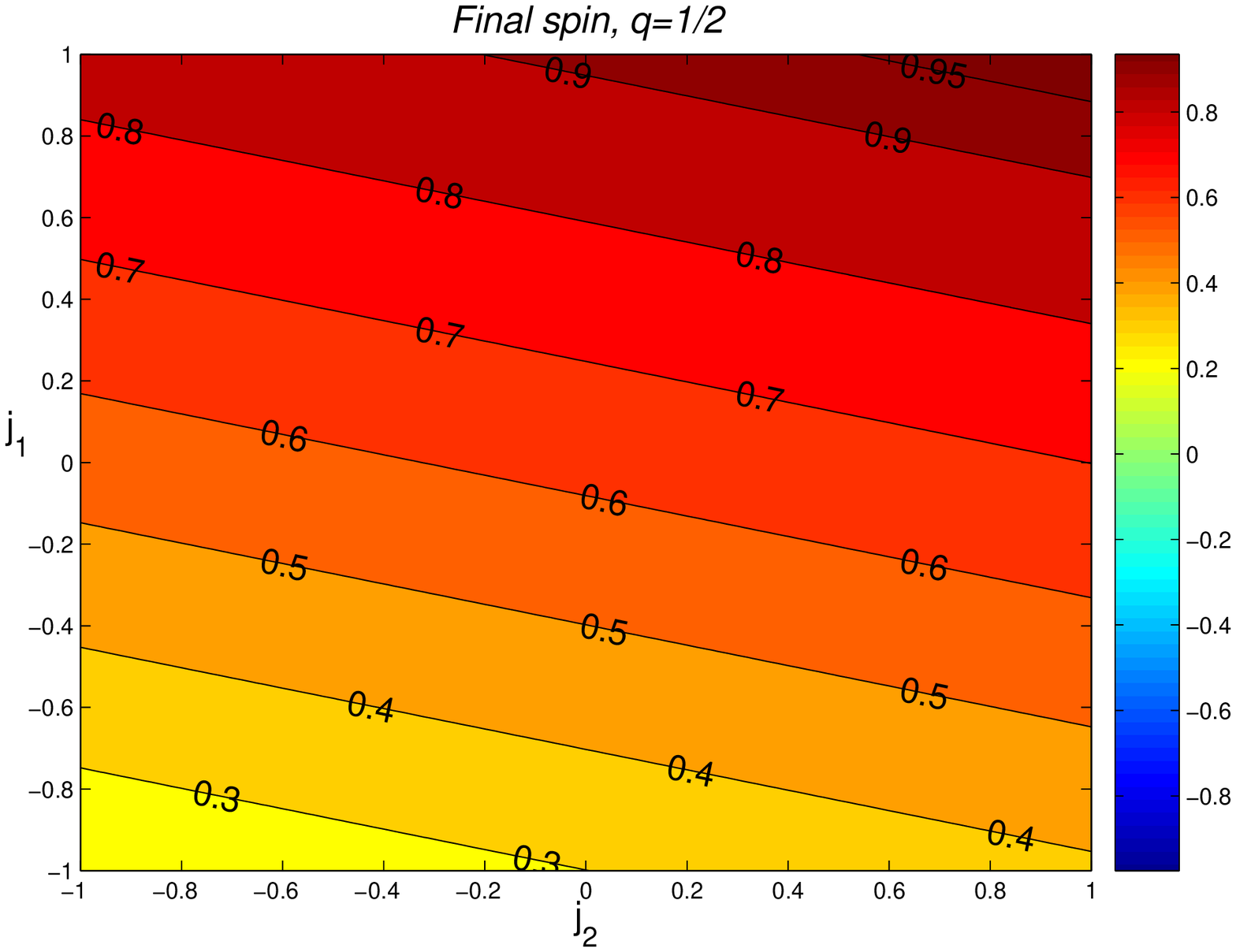}
\includegraphics*[width=0.32\textwidth]{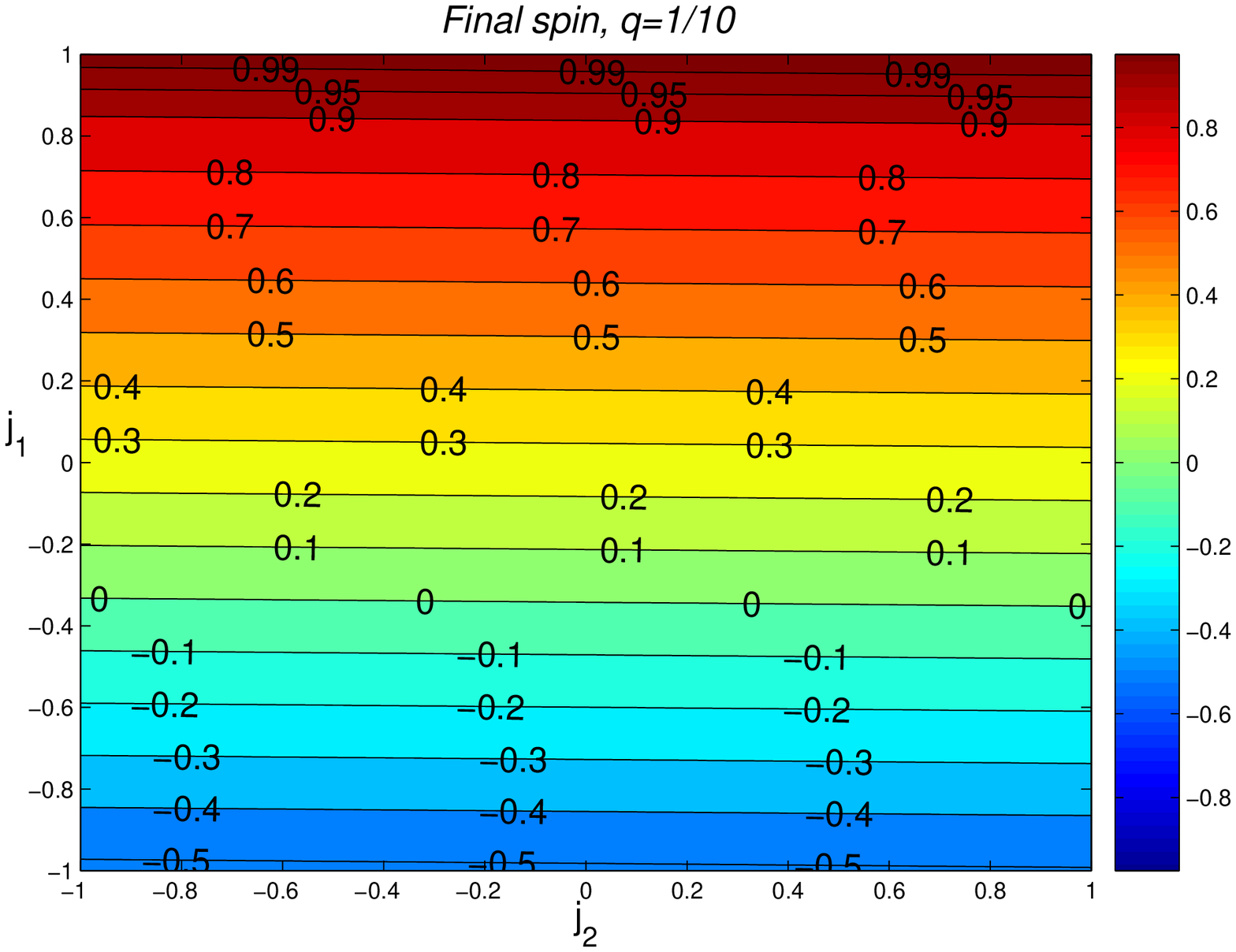}
\includegraphics*[width=0.32\textwidth]{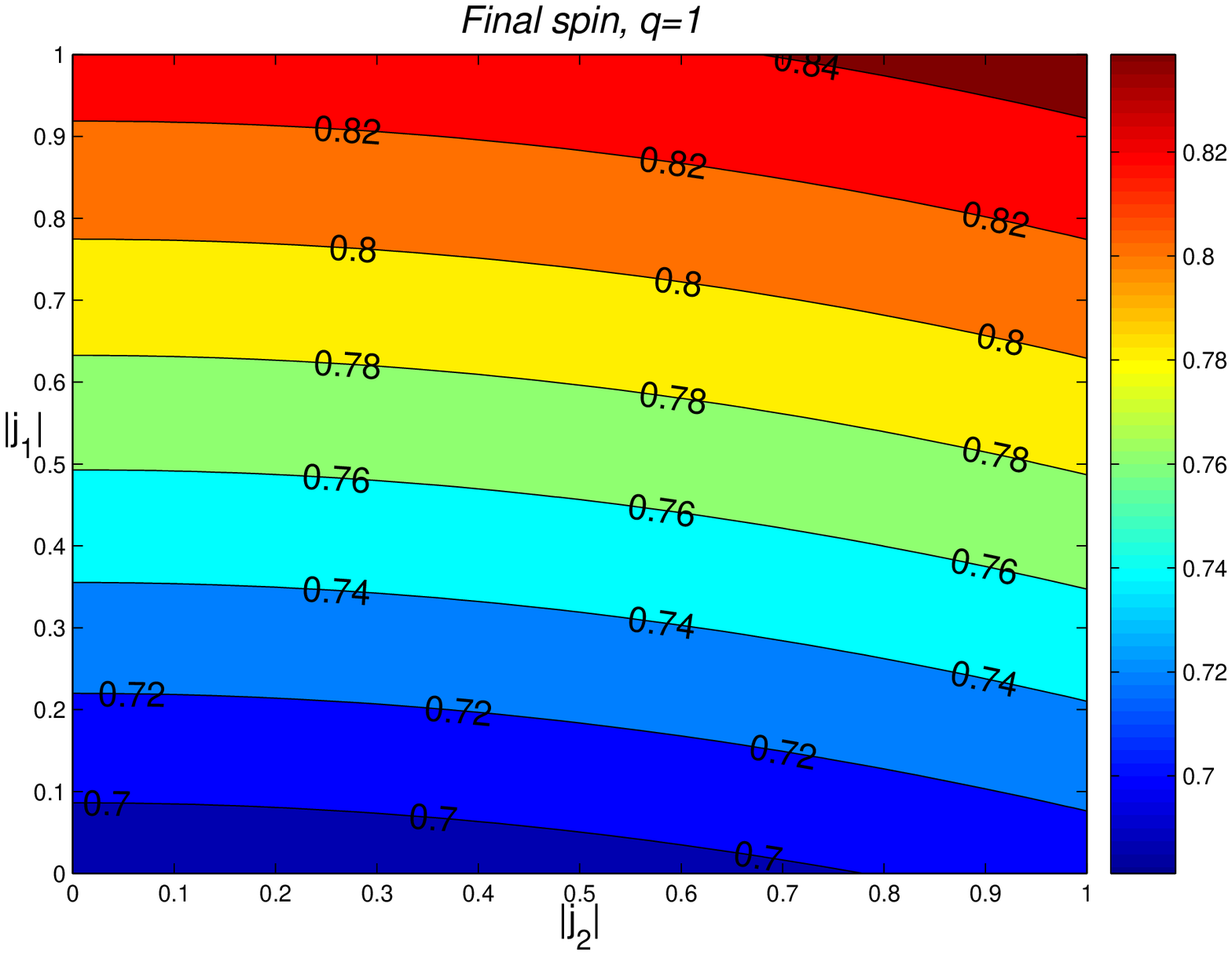}
\hskip 0.2truecm
\includegraphics*[width=0.32\textwidth]{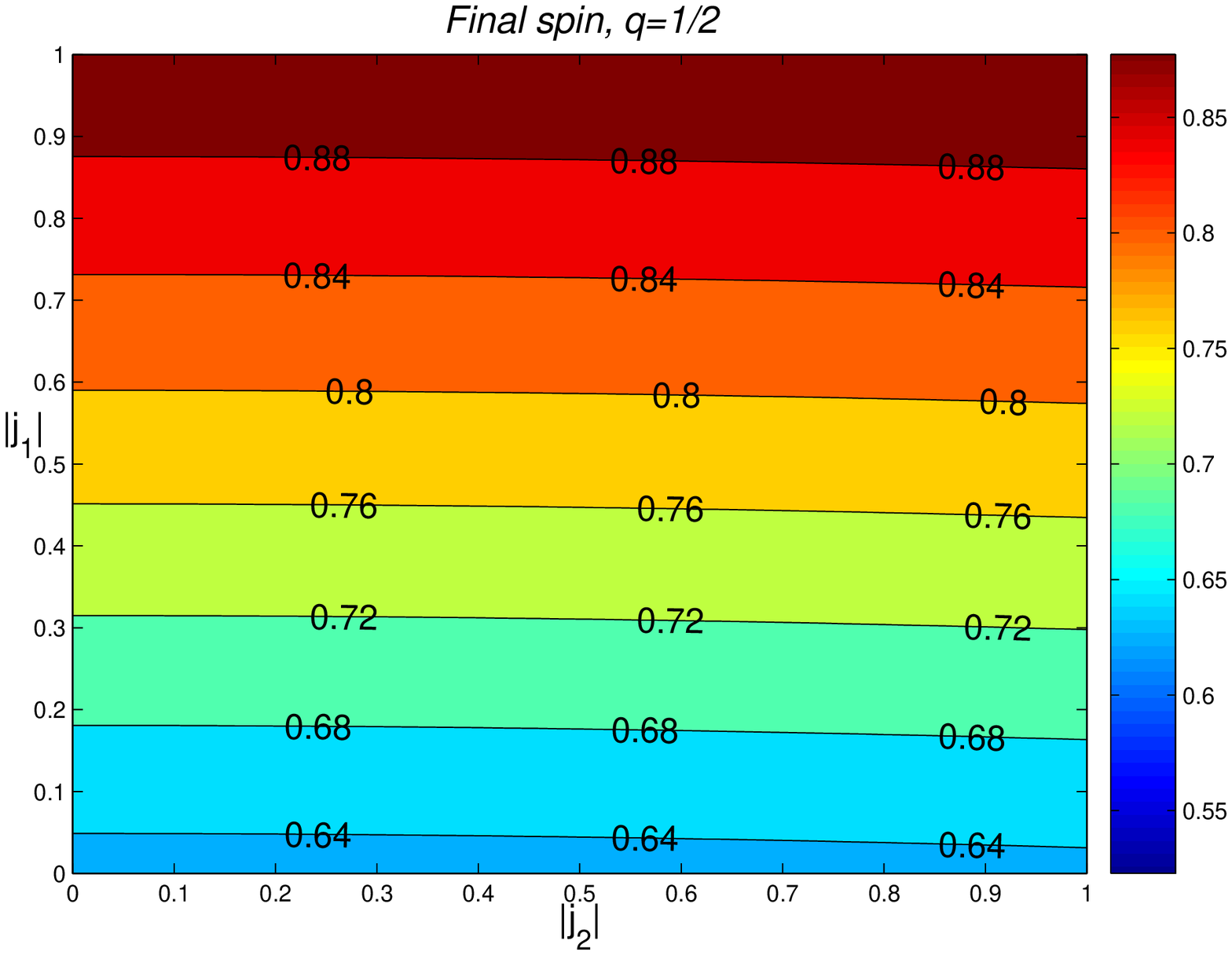}
\hskip 0.2truecm
\includegraphics*[width=0.32\textwidth]{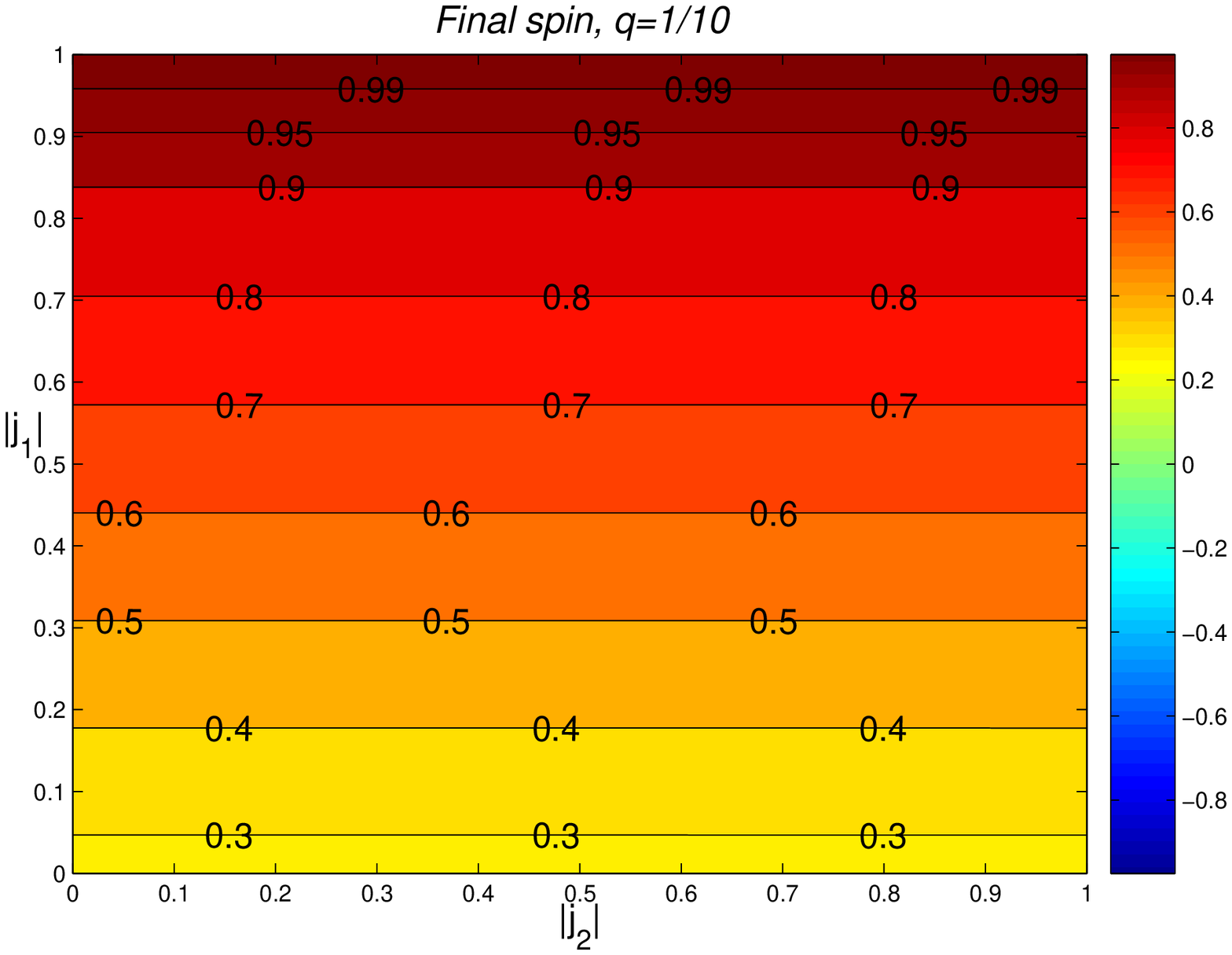}
\caption{\label{fig:contours} Contour plots of the final spin for different
  mass ratios (left to right: $q=1$, $q=1/2$ and $q=1/10$, respectively) in
  the $(\hj{1}\,,\hj{2})$ plane. Recall that $M_2\leq M_1$. We consider three
  different scenarios. In the top row we average over the sky assuming
  isotropy on all three angles. In the middle row we assume that alignment is
  efficient, as proposed by \cite{2007ApJ...661L.147B}. By convention, in this
  case $j_i<0$ means that the spin of BH $i$ is antialigned (rather than
  aligned) with the orbital angular momentum. In the bottom row we set
  $\cos\beta=1$ (so the spin of more massive BH is aligned with the orbital
  angular momentum, and the smaller hole orbits in the equatorial plane of the
  larger) and we assume isotropy in $(\cos\alpha\,,\cos\gamma)$.}
\end{figure*}

\begin{figure}
\vbox{ 
  \centerline{
    \includegraphics*[width=6.6cm]{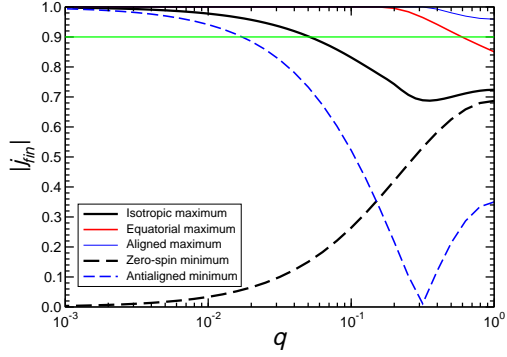}} \figcaption[]
  {\label{fig:spinrange}{\it Solid black (thick) line}: maximum average spin
    in the isotropic case. {\it Solid red (medium thickness) line}: maximum
    average spin for equatorial inspirals. {\it Solid blue (thin) line}:
    maximum spin in the aligned case. {\it Dashed black (thick) line}: minimum
    average spin in the isotropic case.  Since the minimum is attained when
    $\hj{1}=\hj{2}=0$, this line is also the minimum average spin attainable
    by equatorial inspirals, or the minimum spin attainable in the ``aligned''
    case when we rule out the possibility of antialignment. {\it Dashed blue
      (thin) line}: if we do allow for antialignment, the minimum spin can
    become negative (we have a spin flip) when $q\lesssim 0.3$ or so. The
    horizontal (green) line corresponds to a final spin $\hjf=0.9$. When $q$
    is close to one, such large spins are only achievable if alignment is
    efficient.}
}
\end{figure}

We further consider how the dynamics of BH mergers influences the final
spin. The mutual directions of the spins of the holes in a binary with respect
to the orbital plane strongly affect the final spin of the remnant. If the
spins align efficiently with the orbital angular momentum, the spin of the
merger remnant is larger than in the case of random orientations (see Figures
\ref{fig:contours} and \ref{fig:spinrange}). We consider three physically
motivated cases: (1) random orientations (isotropy), which is expected if BH
binaries are in gas-poor host galaxies, or if the orbital plane is not aligned
with the disc plane; (2) alignment between spins and orbital angular momentum,
which \cite{2007ApJ...661L.147B} suggest should be the norm for binaries
forming during gas-rich galaxy mergers; (3) equatorial orbits, where the
smaller BH inspirals in the equatorial plane of the larger. Equatorial mergers
could occur if dynamical friction brings the smaller hole into the plane of a
central disc before it gets into the relativistic regime.

Except in the case of aligned mergers, we find that a sequence of BH mergers
can lead to large spins $\hjf >0.9$ {\it only if} BHs start already with large
spins {\it and} they do not experience many major mergers. This is illustrated
in Figure \ref{fig:spinrange}, where we show that (on average) isotropic
mergers with $q>0.05$ lead to $\hjf < 0.9$. Therefore, the common assumption
that mergers between BHs of similar mass always lead to large spins
\citep[e.g.,][]{1995ApJ...438...62W} needs to be revised. In the isotropic
case, numerical relativity results imply that major mergers tend to produce
BHs with average spins very close to the value $\hjf\simeq 0.69$ resulting
from equal-mass, non-spinning mergers (see again Figure \ref{fig:spinrange}).

\begin{figure*}[thb]
\includegraphics*[width=0.3\textwidth]{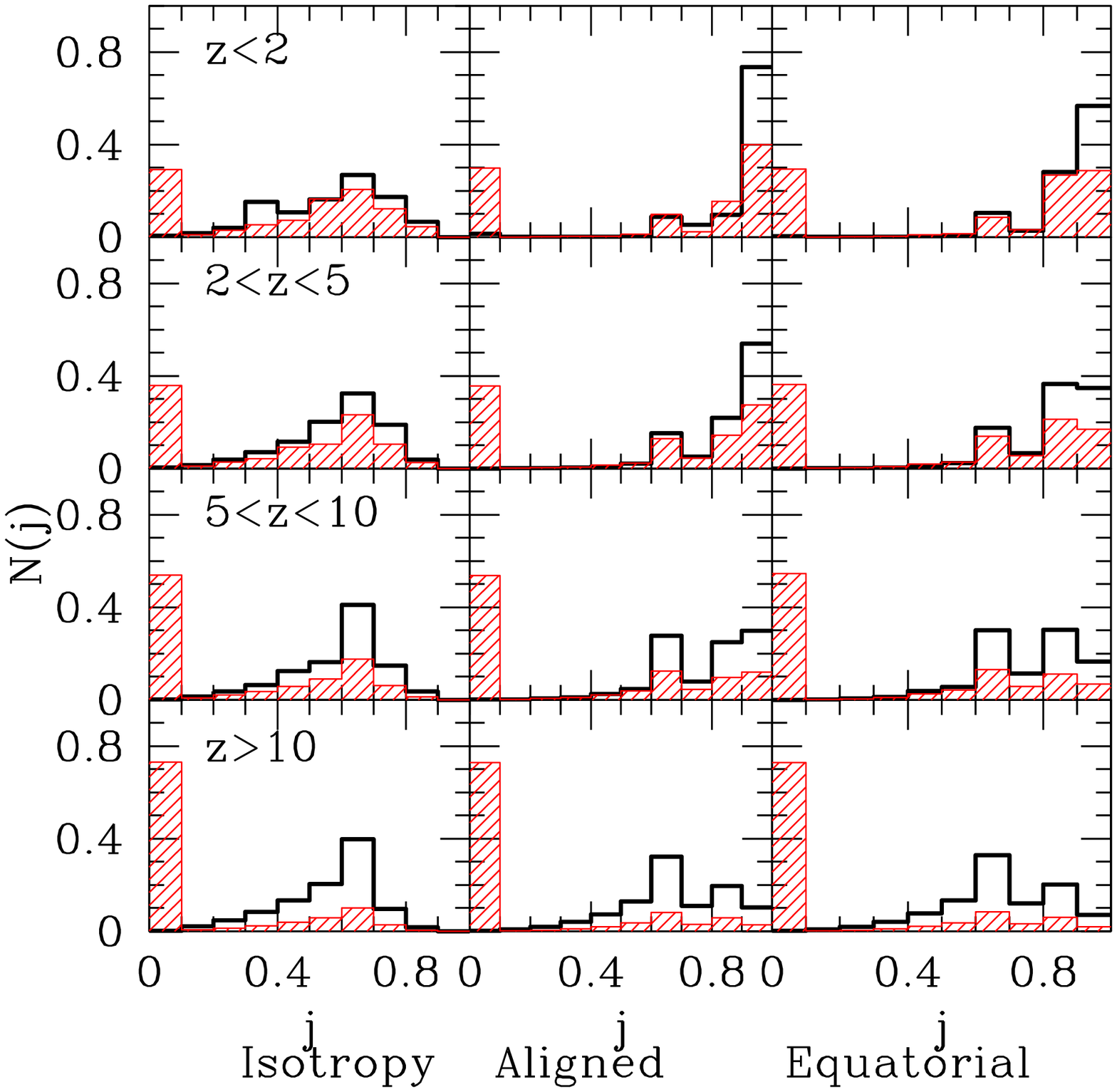}
\includegraphics*[width=0.3\textwidth]{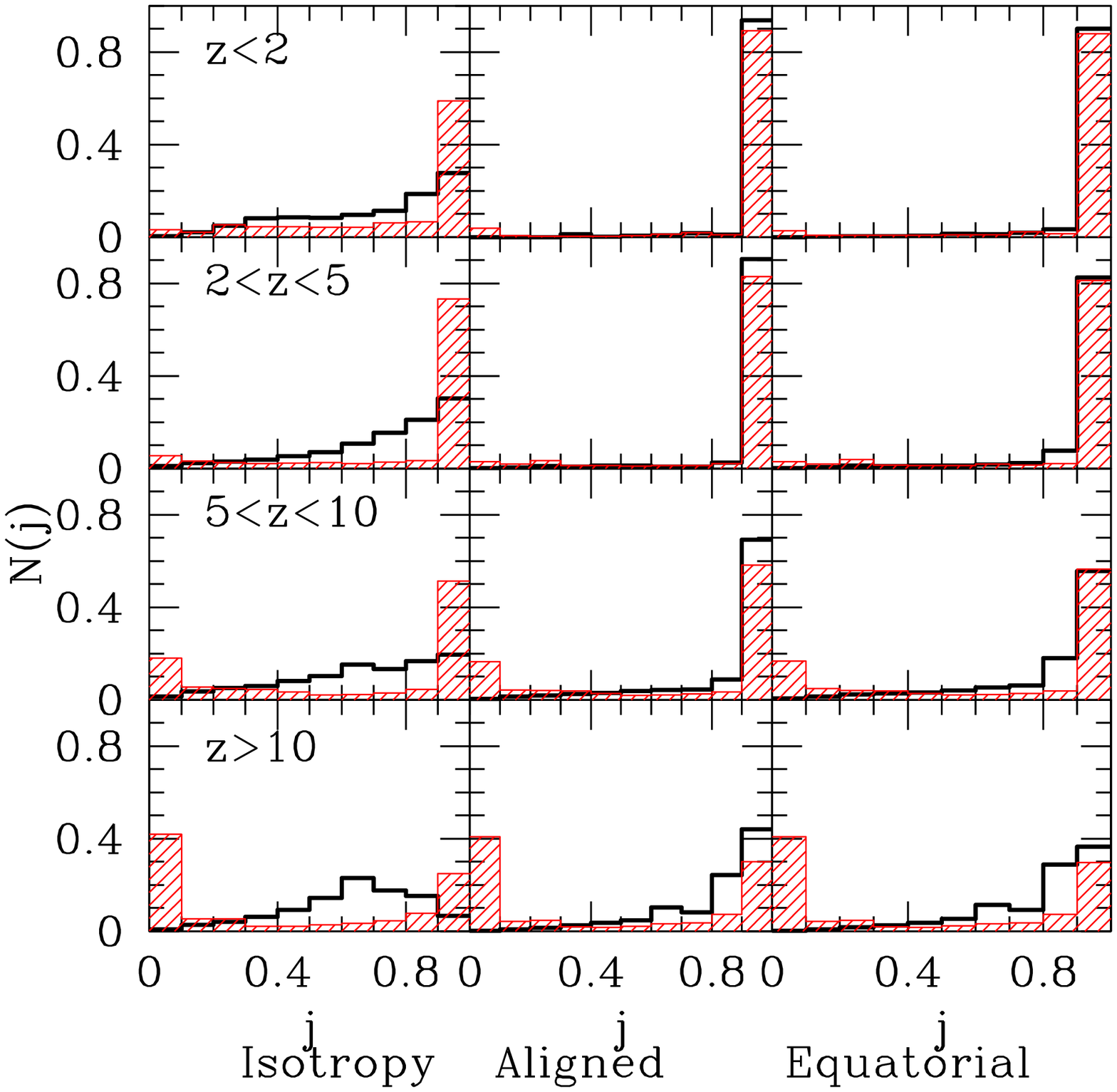}
\includegraphics*[width=0.3\textwidth]{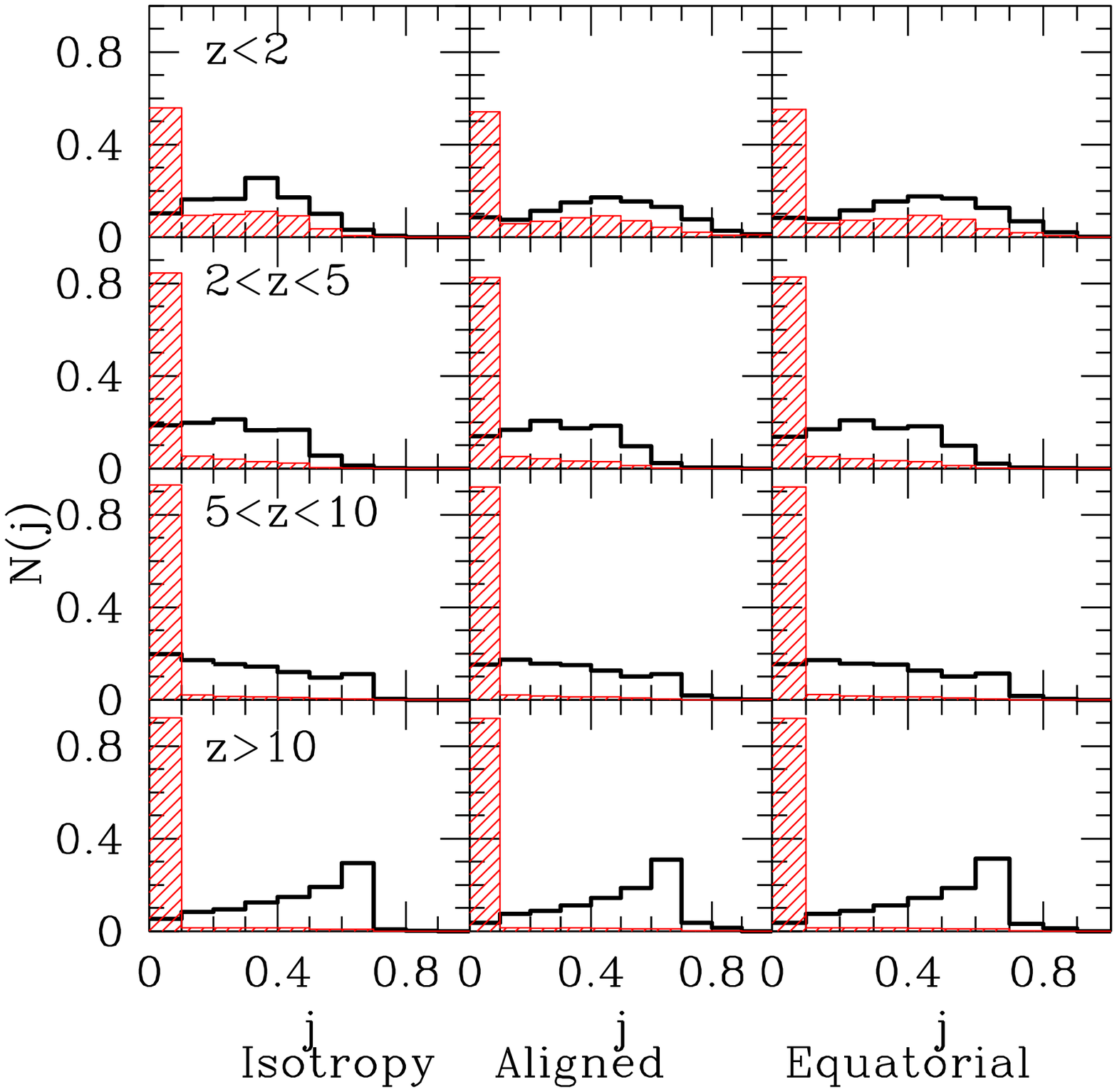}
\caption{\label{fig:mergers-evol} Spin evolution of merging BHs due to:
  mergers only (left); mergers and prolonged accretion (center); mergers and
  chaotic accretion (right). In each plot we consider three representative
  merger scenarios (see text) and we show histograms of the spin distribution
  for different ranges of variability of the redshift $z$. {\it Hatched (red)
    histogram:} spins of the binary members before merger. {\it Thick (black)
    histogram:} spins of BHs after merger.}
\end{figure*}

\begin{figure*}[thb]
\includegraphics*[width=0.3\textwidth]{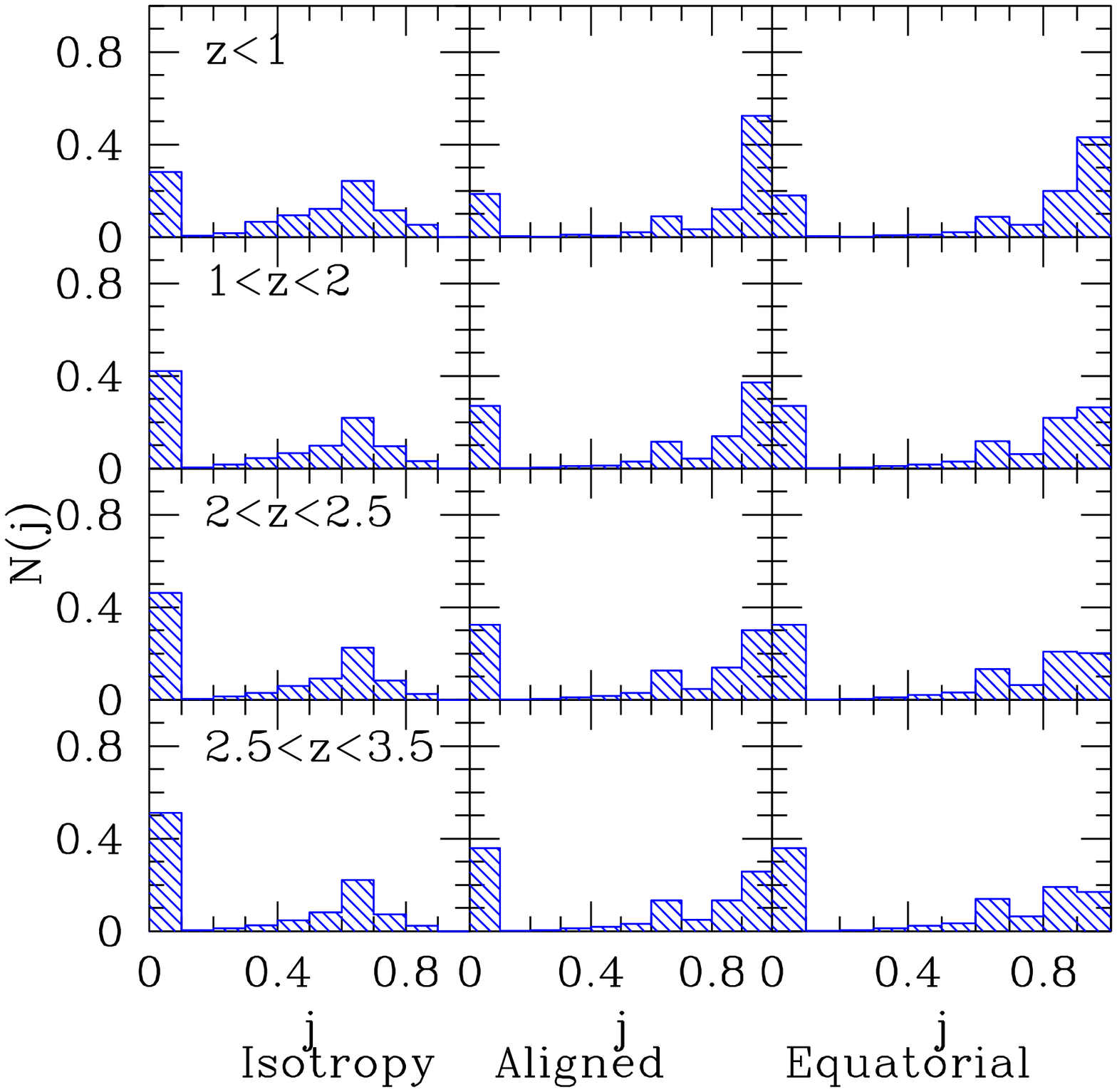}
\includegraphics*[width=0.3\textwidth]{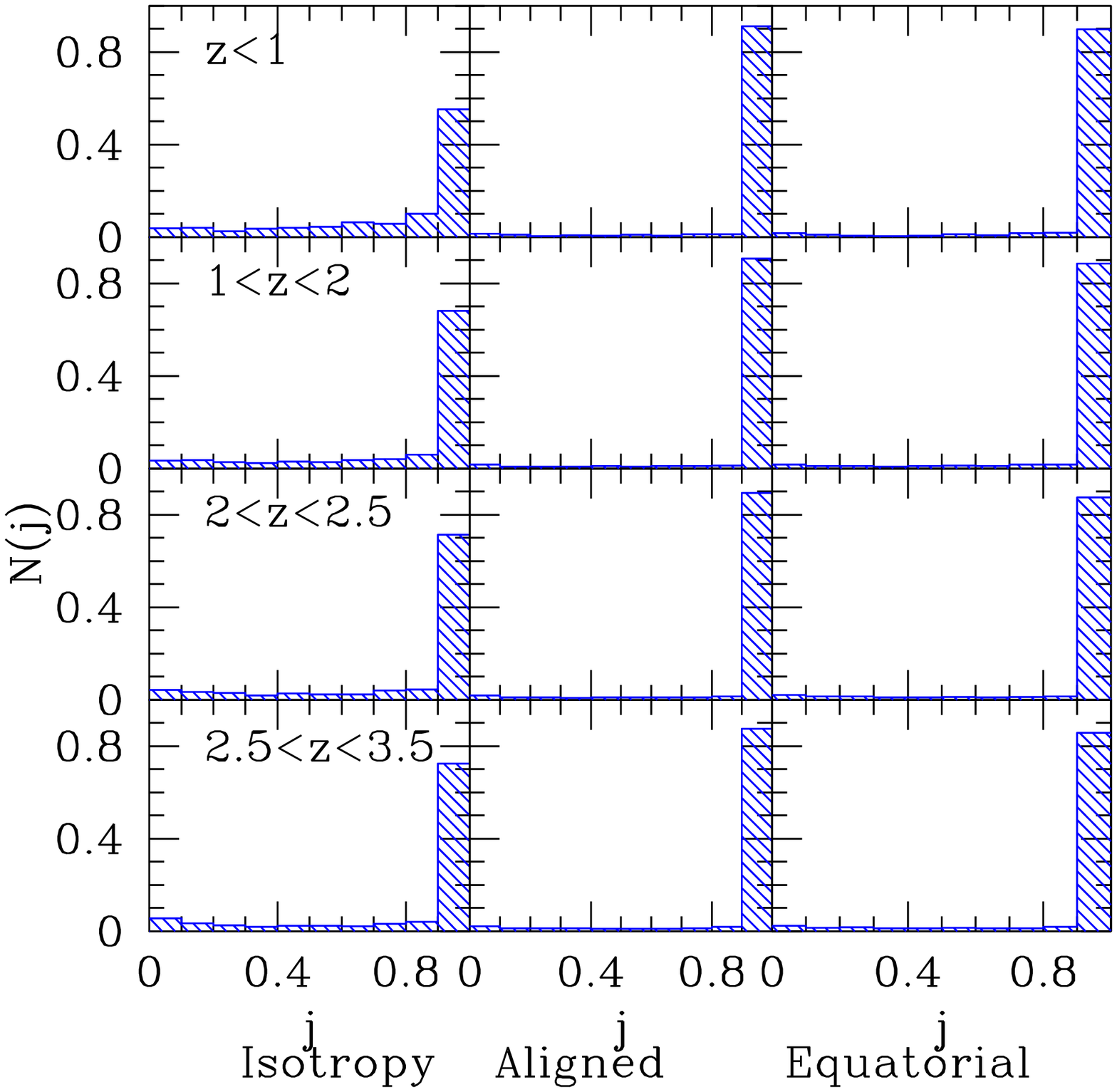}
\includegraphics*[width=0.3\textwidth]{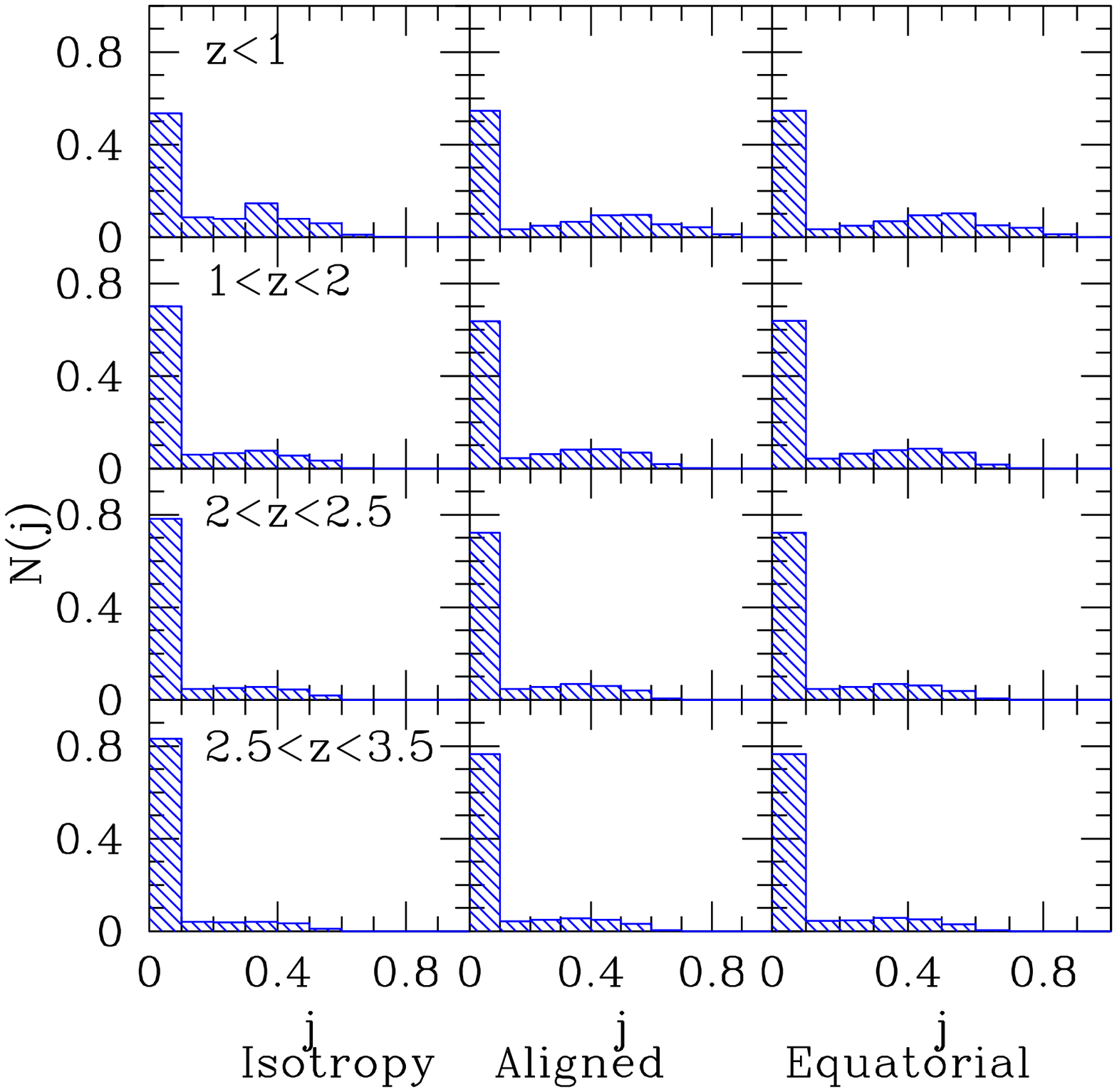}
\caption{\label{fig:all-evol} Spin evolution of all BHs due to mergers only
  (left), mergers plus standard accretion (center) and mergers plus chaotic
  accretion (right).}
\end{figure*}

Our models can be used to test the discriminating power of direct
observational techniques: gravitational-wave measurements and X-ray
spectroscopy. LISA measurements can provide information on the spins of the
two BHs in a binary prior to merger (inspiral) and on the spin of the merger
remnant (ringdown). Such measurements can unequivocally inform us on the
typical spin of the BH population (Figure \ref{fig:mergers-evol}). However, if
LISA measurements of the spin resulting from a binary merger yield $j>0.9$ a
degeneracy remains: a distribution skewed towards large values can occur if
spin is accretion-independent, but alignment is efficient (left plot in Figure
\ref{fig:mergers-evol}), or if instead prolonged accretion is responsible for
spin up (central plot).

This degeneracy is broken by coupling spin measurements of binaries to either
X-ray spectroscopy or EMRIs, as these techniques sample the whole population.
If iron-line measurements and gravitational-wave observations of extreme
mass-ratio inspirals (EMRIs) only yield dimensionless spins $j=J/M^2>0.9$,
then prolonged accretion should be responsible for spin-up, and chaotic
accretion scenarios would be very unlikely.  If instead only a fraction of the
whole population of low-redshift BHs spins rapidly, spin-alignment during
binary mergers (rather than prolonged accretion) could be responsible for
spin-ups.


\acknowledgements We thank Curt Cutler, Tom Prince and the LISA science
performance evaluation taskforce for discussions. E.B.'s research was
supported by an appointment to the NASA Postdoctoral Program at JPL,
administered by Oak Ridge Associated Universities through a contract with
NASA. Copyright 2008 California Institute of Technology. Government
sponsorship acknowledged.


\clearpage

\end{document}